\documentclass[useAMS,usenatbib]{mn2e} \newif\ifdraft \drafttrue
\usepackage{graphicx}
\usepackage{graphics}
\usepackage{rotating}
\usepackage{amsmath}
\usepackage{amssymb}

\newcommand{\vex}{\vspace{1ex}}

\renewcommand{\aa}{A\&A}
\newcommand{\aj}{AJ}
\newcommand{\apj}{ApJ}
\newcommand{\apjs}{ApJS}

\newcommand{\pasp}{PASP}

\newcommand{\mnras}{MNRAS}
\newcommand{\hm}{\hphantom{-}}

\newcommand{\SNR}{{\mbox{\rm SNR}}}

\newcommand{\getlength}[1]{\ifx#1\end \let\next=\relax
            \else\advance\count255 by1 \let\next=\getlength\fi \next}
\newcommand{\ifnularg}[1]{ \count255=0 \getlength#1\end \ifnum\count255=0 }
\newcommand{\ifm}{\makebox{}\ifmmode}
\long\def\ifundefined#1#2#3{\expandafter\ifx\csname
  #1\endcsname\relax#2\else#3\fi}
\newcommand{\beq}   { \begin{eqnarray} }
\newcommand{\eeq}[1]{ \ifnularg{#1} end{eanarray} \else
                      \label{#1}\end{eqnarray}    \fi }
\newcommand{\eeql}   { \end{eqnarray} }
\newcommand{\eeqn}   { \nonumber \end{eqnarray} }

\newcommand{\dss}{\displaystyle}

\newcommand{\ntab}[2]{ \multicolumn{1}{#1}{#2} }

\newcommand{\object}{}
\newcommand{\PIMA}{$\cal P\hspace{-0.067em}I\hspace{-0.067em}M\hspace{-0.067em}A$ }

\newcommand{\Number}[1]{\ifnum#1<10\relax0\number#1\else\number#1\fi}
\newcommand{\isodate}{
\count151=\time
\divide\count151 by 60
\count151=\count151
\multiply\count151 by 60
\count152=\time
\advance\count152 by -\count151
\divide\count151 by 60
\count152=\count151
\multiply\count151 by 60
\count153=\time
\advance\count153 by -\count151
\Number{\year}.\Number{\month}.\Number{\day}--\Number{\count152}:\Number{\count153}
}
\newcommand{\web}[1]{{\sf #1}}

\title[The EVN Galactic Plane Survey --- EGaPS]
      {The EVN Galactic Plane Survey --- EGaPS}
\author[Petrov]{
  \parbox[t]{\textwidth}{
     Leonid Petrov$^{1}$\thanks{E-mail:Leonid.Petrov@lpetov.net}
  }
\vspace{1.0ex} \\
$^{1}$ADNET Systems, Inc./NASA GSFC, Code 610.2, Greenbelt, MD 20771 USA
}

\pagerange{\pageref{firstpage}--\pageref{lastpage}}

\begin{document}

\maketitle
\label{firstpage}

\begin{abstract}

   I present a catalogue of positions and correlated flux densities
of 109 compact extragalactic radio sources in the Galactic plane
determined from analysis of a 48 hour VLBI experiment at 22~GHz
with the European VLBI Network. The median position uncertainty
is 9~mas. The correlated flux densities of detected sources are
in the range of 20 to 300~mJy. In addition to target sources,
nine water masers have been detected, two of them new. I derived
position of masers with accuracies 30 to 200~mas and determined
velocities of maser components and their correlated flux densities.
The catalogue and supporting material is available at
\web{http://astrogeo.org/egaps}.



\end{abstract}


\section{Introduction}

   The differential very long baseline (VLBI) astrometry became
a powerful instrument for a study of the 3D structure and dynamics of
our Galaxy. Analysis of differential observations allows us
to determine parallaxes of compact radio sources with accuracies down
to several tens of microarcseconds, positions with accuracies
0.05~mas and proper motions. The differential astrometry observations
are made in the so-called phase referencing mode when radio
telescopes of a VLBI network rapidly switch from a target source
to a near-by calibrator. Analysis of differential fringe phases allows
to dilute the contribution of mismodeled path delays in the atmosphere
by the factor of target to calibrator separation in radians, i.e.
20--50 times. Nowadays, observations in the phase-referencing mode
are widely used for imaging faint sources. The atmosphere path delay
fluctuations limit the integration time, depending on frequency,
to 0.2--10 minutes. Phase  referencing observations allow to overcome
this limit and to integrate the signal for hours. According to
\citet{r:wrobel_pr}, 63\% of Very Long Baseline Array (VLBA)
observations in 2003--2008 were made in the phase referencing mode.

   However, the feasibility of phase-referencing observations is
determined by the availability of a pool of calibrators with
precisely known positions. The first catalogue of source coordinates
determined with VLBI contained 35~objects \citep{r:first-cat}.
Since then, hundreds of sources have been observed under geodesy
and astrometry VLBI observing programs at 8.6 and 2.3~GHz
(X and S bands) using the Mark3 recording system at the International
VLBI Service for Geodesy and Astrometry (IVS) network. Analysis
of observations in 1980s and 1990s resulted in the ICRF catalogue
of 608~sources \citep{r:icrf98}. Later, using the VLBA, positions
of over 5000 other compact radio sources were determined in the
framework of the VLBA Calibrator Survey (VCS)
\citep{r:vcs1,r:vcs2,r:vcs3,r:vcs4,r:vcs5,r:vcs6}, the VIPS program
\citep{r:vips,r:astro_vips}, the BeSSel Calibrator
Survey \citep{r:br145a}, and in the number of on-going programs:
the geodetic VLBA program RDV \citep{r:rdv}, the program of a study
of active galaxy nuclea (AGNs) at parsec scales detected with {\it Fermi}
(Kovalev \& Petrov, paper in preparation), the program of observing
radio-loud 2MASS galaxies (Petrov, paper in preparation), and
the program of observing optically bright quasars
\citep{r:bou08,r:bou11,r:obrs1}. The use of the Australian Long
Baseline Array (LBA) extended the catalogue of calibrators to
the zone $[-90\degr, -40\degr]$ \citep{r:lcs1}.

  By April 2011, the total number of radio sources with positions
determined with the VLBI in the absolute astrometry mode reached 6123
and it continues to grow. {\it On average}, the probability of finding
a calibrator within a $3\degr$ radius of a given position is 97\%.
However, the density of calibrators in the Galactic plane is still
low and an increase of the number of calibrators is badly needed
for many VLBI Galactic astronomy projects. Finding extragalactic sources
visible through the Galactic plane region is more difficult for several
reasons. First, the region is crowded with many galactic objects. Combining
independent low-resolution observations at different frequencies
makes determination of their spectra problematic due to a risk
of source mis-identification. Second, many potential candidates with
flat spectra are extended galactic objects, such as planetary nebulae
or compact HII regions, that cannot be detected with VLBI. Third,
the apparent angular size of extragalactic objects observed through
high plasma density near the Galactic plane is broadened by Galactic
scattering and cannot be detected at low frequencies on baselines
longer than several thousand kilometers.

  To address the problem of increasing the density of calibrators
within the Galactic plane, we made in 2005--2006 a dedicated blind
fringe survey of 2496 objects with the VERA \citep{r:vera_fss} radio
interferometer at the K-band (22~GHz). Those sources which were
detected with the VERA were re-observed with the VLBA in 2006
at K-band in the framework of the VGaPS campaign \citep{r:vgaps}.
These observations allowed to determine positions of 176 new sources
within 6\degr of the Galactic plane with the median accuracy of
0.9~mas. The detection limit of VERA was 200--300~mJy, the detection
limit of the follow-up VLBA observations was 70--90~mJy depending
on a baseline\footnote{The K-band receivers at the VLBA stations were
upgraded in 2007, after these observations, which improved the
sensitivity of the array at 24~GHz at the same recording rate
by a factor of 2 according to \citet{r:cwa08}.}. Independently,
a team led by Mark Reid searched for Galactic plane calibrators in 2010
using the VLBA in the framework of BeSSeL project and reported
detection of 198 sources \citet{r:br145a}, 82 of them are new.

  In this paper I present results from a 48~hour experiment
at the European VLBI Network (EVN) observed in October 2009 called
the EVN Galactic Plane Survey (EGaPS). The goal of this experiment
is to further increase the calibrator source density in the zone
within 6\degr of the Galactic plane and with declinations
$>-20\degr$. Using highly sensitive antennas, I was able to detect
sources as weak as 20~mJy, a factor of 4 weaker than in the prior
VGaPS campaign. These calibrators are aimed to be used for Galactic
astronomy projects. The selection of candidate sources and
the scheduling strategy is discussed in section~\ref{s:selection}.
The station setups during the observing sessions is described
in section~\ref{s:observations}. The correlation and post-correlation
analysis is discussed in section \ref{s:analysis}. The catalogue
of source positions and correlated flux densities is presented
in section~\ref{s:catalog}. An error analysis of observations,
including evaluation of systematic errors, is given in
subsection~\ref{s:errors}, and the results are summarized
in section~\ref{s:summary}.

\section{Candidate source selection}
\label{s:selection}

  In the past, the list of bright compact sources detected at the
Very Large Array (VLA) and/or MERLIN served as a pool for candidates
for VLBI calibrator surveys. Almost all these objects have been
already surveyed with the VLBA. Since the combined catalogue of
sources observed in the absolute astrometry mode is complete at
the 150--200~mJy level, new calibrators should be relatively faint
objects. Most all-sky surveys at frequencies higher than 2~GHz are
either incomplete and/or not deep enough. As a result, information
about spectral indices is sparse and is often unreliable.
For this reason, it is more difficult to find remaining good flat
spectrum candidate sources.

  New calibrator candidates were found by analyzing 395 radio
astronomical catalogues using the CATS database \citep{r:cats}.
The database includes NVSS \citep{r:nvss}, CLASS \citep{r:class},
CRATES \citep{r:crates}, ATCA survey at 20~GHz \citep{r:at20g}, and
many other catalogues. Initially, I selected 1100~sources which
satisfy the following criteria:
\begin{itemize}
   \item were not observed previously with VLBI;
   \item have galactic latitude $|b|<6^\circ$;
   \item have declination $>-20\degr$;
   \item have at least two measurements of their flux density that
         allow estimation of their spectral indices at frequencies
         higher than 2~GHz;
   \item have single-dish flux densities extrapolated to 22~GHz
         $>$ 80mJy;
   \item have spectral indices $\alpha$ flatter than $-0.5$
         ($S\sim\nu^\alpha$).
\end{itemize}

  The majority of these sources were selected by
cross-identification of the NVSS and GB6 \citep{r:gb6} catalogues.
Then I scrutinized the source spectra and rejected approximately
50\% of the sources with unreliable spectra or the objects known
as galactic sources. The remaining list contains 559 candidates.

\subsection{Observation scheduling}

\begin{figure}
  \caption{The EVN stations that participated in the EGaPS experiment.
           The longest meridional baselines {\sc noto}/{\sc onsala60}
           is 2280~km long. The longest longitudinal baseline
           {\sc noto}/{\sc yebes40m} is 1616~km long.}
  \ifdraft
     \par\vspace{-2ex}\par
     \includegraphics[width=0.48\textwidth]{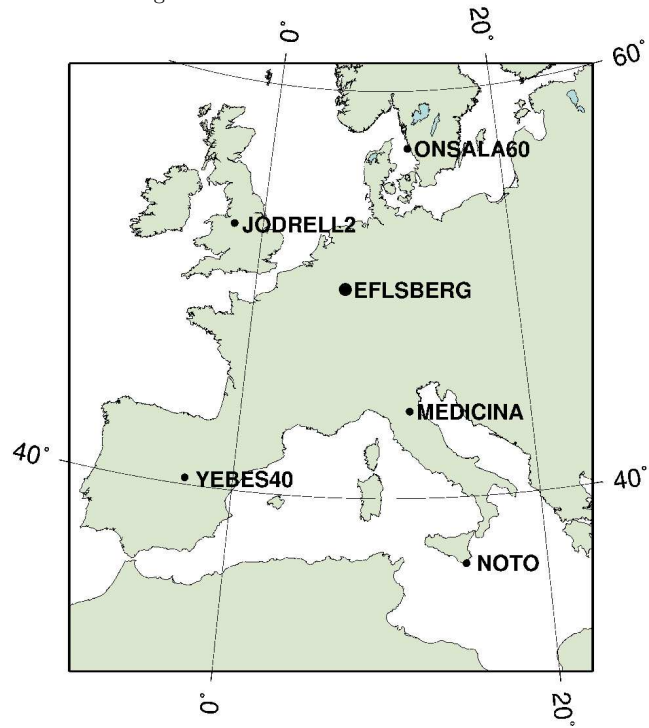}
     \par\vspace{-3ex}\par
  \else
     \includegraphics[width=0.48\textwidth]{evn_map_1200.eps}
  \fi
  \label{f:evn}
\end{figure}

  The following stations were scheduled for 48 hour observations:
{\sc eflsberg}, {\sc jodrell2}, {\sc medicina}, {\sc noto},
{\sc onsala60}, and {\sc yebes40m} (see Figure~\ref{f:evn}.
The observation schedule was prepared with the software
program {\sf sur\_sked}. The scheduling goal was to observe each
target source at all antennas of the array in 2~scans of
120~seconds each.

  The scheduling algorithm found the sequence of sources that minimized
slewing time. The minimum time between consecutive observations of the
same source was set to 4~hours. For deciding which source to put into
the schedule, the algorithm calculated the elevation of each candidate
object, the slewing time, the time interval between previous
observation of the same source, and the ratio of the remaining visibility
time to the total visibility time. For each source with the elevation
exceeding 15\degr above the horizon, a score was computed. The score
is the function of slewing time, the ratio of the remaining visibility
time to the total visibility time, and the time interval from the
previous observation. A source with the highest score was put into
the schedule and then the procedure was repeated.

  Every 1.5~hr a burst of 4 strong compact sources with known maps
from the K/Q survey \citep{r:kq} was observed: two strong objects
at elevation angles 10--$30^\circ$  and two strong objects
at elevations 50--$90^\circ$. The purpose of scheduling these
calibrators was a)~to allow estimation of the troposphere path delay
in zenith direction; b)~to evaluate the atmosphere opacity; c)~to
use them for complex bandpass calibration; d)~to provide a strong
connection between the new catalogue to the old catalogue of
compact sources.

  The schedule had 369 target objects and 69 calibrators. Among
target sources, 344 objects were observed in  2 scans and 25 sources
were observed in 1 scan.

\begin{figure*}
  \caption{The distribution of calibrator sources within 10\degr
           of the Galactic plane. The filled disks denote sources
           detected in the EGaPS campaign. The hollow circles
           denote calibrator sources known from previous observations.
           New sources from the BeSSeL VLBI survey are not shown.}
  \ifdraft
     \includegraphics[width=1.00\textwidth,clip]{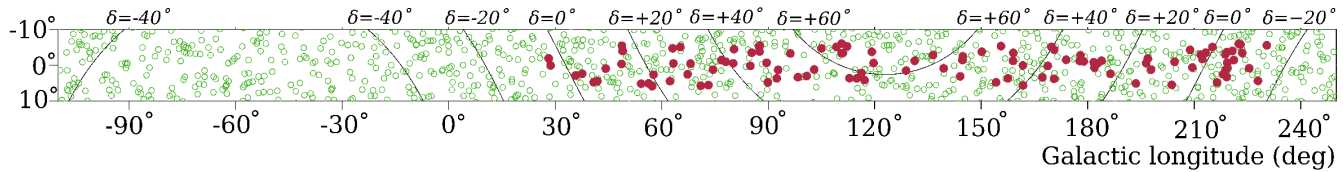}
     \par\vspace{-2ex}\par
  \else
     \includegraphics[width=1.00\textwidth,clip]{distr_1200.eps}
  \fi
  \label{f:distr}
\end{figure*}

\section{Observations}
\label{s:observations}

  Observations took place on 2009 October 27--29. Eight intermediate
frequencies (IF), both upper and lower sub-band in the range of
[22.09999,  22.35599]~GHz were observed in a single left circular
polarization with two bits per sample. The data were recorded with
the Mark-5 system with the aggregate rate 1024~Mbit/sec.
The contiguous frequency setup is far from optimal for an absolute
astrometry experiment that utilizes group delays, since the uncertainties
of a group delay at a given signal-to-noise ratio (\SNR) is reciprocal
to the root mean squares (rms) of IFs. Two stations, {\sc medicina}
and {\sc onsala60} could record within 720~MHz, three stations
equipped with the VLBA data acquisition system, {\sc eflsberg}
and {\sc noto} {\sc yebes40m}, could record within 500~MHz and
{\sc jodrell2} could record within 160~MHz. Spreading IFs over 500~MHz
would reduce group delay uncertainty by a factor of 2.4 at all
baselines, except {\sc jodrell2} where it would increase the
uncertainty by a factor of 2 since some channels will be missing.
However, in order to record a band wider than 256~MHz, the local
oscillator (LO) frequency should be changed with respect to those that
are used in other EVN experiments. This would require manual intervention.
When I realized that for logistical reasons there is a high risk that
this change would not be made at some stations, which could ruin
the experiment, I fell back to the contiguous frequency setup.

  Station {\sc yebes40m} lost first 7 hours of the experiment
because of the Mark-5B failure. The system temperature of
{\sc yebes40m} was around 210~K during next three hours because the
vertex was closed. It dropped to the normal range of 60-100~K
when the operator noticed it and opened the vertex. Station
{\sc jodrell2} did not show fringes in IF11--IF16.

\section{Data analysis}
\label{s:analysis}

  Analysis of the EGaPS data is similar to that performed for
the VLBA Galactic plane calibrator survey. Detailed description
of the analysis technique can be found in \citet{r:vgaps}.
Here the analysis procedure is briefly outlined.

\subsection{Data correlation and post-correlation analysis}

  The data were correlated at the Bonn DiFX software correlator
\cite{r:difx} with the accumulation periods of 0.12~s and with
the spectral resolution of 125~KHz. These correlation parameters
provided the field of view over $1'$, comparable with the width
of the beam of {\sc eflsberg} radio telescope. I anticipated
the a~priori source coordinates of some sources may be wrong
at the arc-minute level due to source mis-identification.

  The correlator generates the spectrum of the cross-correlation
function that was written in the format compliant with the FITS-IDI
specifications \citep{r:fits-idi}. Data analysis was performed with
software \PIMA$\!$\footnote{Available at {\tt http://astrogeo.org/pima}.}.
At the first step, the fringe amplitudes were corrected for the
signal distortion in the sampler. Observations of strong sources were
used for deriving complex bandpasses. Since in 2010 the DiFX correlator
did not have an ability to extract phase calibration signal from
the data, no phase calibration was applied. After that, the group
delay, phase delay rate, group delay rate, and fringe phase were
determined for each scan at each baseline using the fringe fitting
procedure. These estimates maximize the sum  of the cross-correlation
spectrum coherently averaged over all accumulation periods of a scan
and over all frequency channels in all IFs. After the first run of fringe
fitting, 8 observations at each baseline with the strongest \SNR\ were
used to adjust the station-based complex bandpass corrections,
and the procedure of computing group delays was repeated with
the bandpass applied.

   Then the results of fringe fitting were exported to the
VTD/post-Solve VLBI analysis software\footnote{Available at
{\tt http://astrogeo.org/vtd}.} for interactive processing group
delays with the \SNR\ high enough to ensure the probability of false
detection less than 0.001. Analysis of the probability distribution
function of achieved \SNR\ allowed me to find that the probability
of a false detection in our experiment is less than 0.001
when \SNR $>6.14$. Theoretical path delays were computed
according to the state-of-the art parametric model as well as
their partial derivatives. The model included the contribution of the
ionosphere to group delay and phase delay rate computed using the
total electron contents maps derived from analysis of Global
Positioning System (GPS) observations by the analysis center
CODE \citep{r:scha98}. Small differences between group delays
and theoretical path delay as well as between the measured and
theoretical delay rate were used for an interactive estimation
of corrections to a parametric model that describes the observations
with least squares (LSQ). Coordinates of target sources, positions
of all stations, except the reference one, parameters of the splines
that describe corrections to the a~priori path delay in the neutral
atmosphere in the zenith direction for all stations, and parameters
of another splines that describe the clock function were solved
for in a single LSQ solution using group delays. Outliers were
identified and temporarily suppressed, and additive corrections
to weights of observables were determined. Then the fringe fitting
procedure was repeated for outliers with phase delay rates and
group delays evaluated in a narrow window around the expected value
computed on the basis of results of the previous LSQ solution.
New estimates of group delays for points with probabilities of false
detection less than 0.1, which corresponds to the \SNR $> 4.9$ for the
narrow fringe search window, were used in the next step of the interactive
analysis procedure. For observations detected with the narrow
window search the status outlier was lifted and they were selected back
for using in further analysis. Parameter estimation, elimination of
remaining outliers and adjustments of additive weight corrections was then
repeated. In total, group delays of 2471 observations out of 11675
scheduled were used in the solution. There were 178 sources detected
in 3 to 28 observations, including 109 target sources and
69 calibrators.

\subsection{Source position determination}

  Results of the interactive solution provided a clean dataset of
group delays with updated weights from the EGaPS experiment. The dataset
that was used for the final parameter estimation utilized all dual-band
S/X data acquired under absolute astrometry and space geodesy programs from
April 1980 through February 2011 and included the K-band data from the EGaPS
experiment, in total 8~million observations. The estimated parameters
are right ascensions and declination of all sources, coordinates
and velocities of all stations, coefficients of B-spline expansion
of non-linear motion for 16 stations, coefficients of harmonic site
position variations of 48 stations at four frequencies: annual,
semi-annual, diurnal, semi-diurnal, and axis offsets for 67 stations.
They were adjusted using all the data. Estimated parameters also included
Earth orientation parameters for each observing session, parameters
of clock function and residual atmosphere path delays in the zenith
direction modeled with the linear B-spline with interval
60 and 20 minutes respectively. All parameters were estimated
in a single LSQ run.

  The system of LSQ equations has an incomplete rank and defines
a family of solutions. In order to pick a specific element from this
family of solutions, I applied the no-net rotation constraints on
the positions of 212~sources marked as ``defining'' in the ICRF
catalogue \citep{r:icrf98} that required the positions of these source
in the new catalogue to have no rotation with respect to the position
in the ICRF catalogue. No-net rotation and no-net-translation
constraints on site positions and linear velocities were also applied.
The specific choice of identifying constraints was made to preserve
the continuity of the new catalogue with other VLBI solutions made
during last 15 years.

  The global solution sets the orientation of the array with respect
to the ensemble of $>5000$ extragalactic remote radio sources.
The orientation is defined by the continuous series of Earth
orientation parameters and parameters of the empirical model of
site position variations over 30 years evaluated together with
source coordinates. Common sources observed in the EGaPS experiment
as amplitude calibrators provided a connection between the new
catalogue and the old catalogue of compact sources.

  A valuable by-product of analysis of EGaPS observations is
estimates of {\sc jodrell2} positions (see~Table~\ref{t:jod}). This
is the second experiment with participation of this station under
astrometry or geodesy programs. The previous EVN  experiment TP001
carried out on 2000 November 23 was made at 5~GHz with the bandwidth
spread over 108~MHz \citep{r:cha02}. Comparison of results of reanalysis
of TP001 and EGaPS made using exactly the same a~priori model and
parameter estimation technique showed that the differences between
{\sc jodrell2} position reduced to the J2000.0 epoch using the
velocities predicted by the NNR-NUVEL1A plate tectonic model
\citep{r:nuvel} are 38, 20, and 9~mm at X, Y and Z coordinate
respectively. Although position of {\sc jodrell2} from the 5~GHz
observations with formal uncertainties 10--20~mm suffered from
residual errors in modeling path delay through the ionosphere which
are 20 times greater than at 22~GHz, the differences in positions
are within 1--2 reported uncertainties.

\begin{table}
   \caption{Estimates of {\sc jodrell2} positions in meters on
            epoch 2009.10.27}
   \label{t:jod}
   \begin{tabular}{cr@{\:\:}c@{\:\:}l}
      X  & 3822846.633 & $\pm$ & 0.026  \\  
      Y  & -153802.071 & $\pm$ & 0.009  \\  
      Z  & 5086286.064 & $\pm$ & 0.036  \\  
   \end{tabular}
\end{table}

\subsection{Position error analysis}
\label{s:errors}

  The formal uncertainties of semi-major axis of the error ellipses
of position estimates of 109 target sources range from 2 to 60~mas
with the median value of 8~mas. They are based on the error
propagation law of group delay uncertainties derived by the fringe
fitting procedure.

  Including in observing schedule a considerable number of calibrator
sources with positions known at the sub-mas level facilitated the error
analysis. In order to evaluate the level of systematic errors,
I used a similar technique that was developed for processing VGaPS
observations. The list 69 calibrators was ordered in increasing their
right ascensions and split it into two subsets with even and
odd indices. I made two special global solutions using all the data
from 1980 through 2011. I excluded in solution A calibrators with
even indices from all experiments, but the EGaPS. I excluded
in solution B calibrators with odd indices from all experiments, but the
EGaPS. I compared positions of calibrator sources from solutions
A and B with their position from the dual-band global solution C
that used all sources in all experiments, except the EGaPS.
Positions of sources from solution C is determined with accuracies
0.1--0.3~mas and can be considered as true for the purpose of this
comparison.

  The wrms of the differences of 69 calibrator sources from special
solutions A and B and the reference solution C are 1.8~mas in
declinations and 3.7~mas in right ascensions scaled by
$\cos\delta$ with $\chi^2$ per degree of freedom 1.3 and 3.3
respectively. If to add in quadrature to uncertainties in right ascensions
and declinations the noise with the standard deviations
4.0~mas$/cos \delta$ and 1.0~mas respectively, the $\chi^2$
per degree becomes close to 1. Therefore, I inflated the reported
uncertainties in right ascensions and declinations of target sources by
4.0~mas$/cos \delta$ and 1.0~mas in quadrature.

\subsection{Data analysis: correlated flux density determination}

  Each detected source has from 3 to 30 observations, with the median
number of 8. The dataset is too sparse to produce meaningful images.
In this study I limited my analysis with mean correlated flux
density estimates in two ranges of lengths of the baseline projections
onto the plane tangential to the source without inversion of calibrated
visibility data. Information about the correlated flux density is needed
for evaluation of the required integration time when an object is used
as a phase calibrator.

  Analysis of system temperature measurements revealed variations with
time and with elevation angle. The measured system temperature is
considered as a sum of two terms: the receiver temperature $T_{r}$ and
the contribution of the atmosphere:
\begin{eqnarray}
   T_{\rm sys} = T_{r} \: + \: T_{\rm atm} [ 1 - e^{-\beta \, m(e)} ] ,
   \label{e:e3}
\end{eqnarray}
   where $T_{atm}$ is the average temperature of the atmosphere,
$\beta$ is the atmosphere opacity, and $m(e)$ is the wet mapping
function: the ratio of the neutral atmosphere non-hydrostatic path
delay at the elevation $e$ to the atmosphere non-hydrostatic path
delay in the zenith direction. I omitted in expression \ref{e:e3}
the ground spillover term that was not determined for these antennas.
Both, the receiver temperature and the atmosphere opacity,
depend of time. Since in astrometric analysis I estimated for each station
the non-hydrostatic component of the atmosphere path delay in the zenith
direction which is closely related to the integrated column of
the water vapor, I can use these estimates for modeling time dependence
of the opacity. I present the system temperature as
\begin{eqnarray}
   T_{\rm sys} = \dss\sum_i T_{ri} B^1_i(t) \: + \: T_{\rm atm}
             [ 1 - e^{-(a + b\tau_{\rm atm}(t)) \, m(e)} ] ,
   \label{e:e4}
\end{eqnarray}
  where $a$ and $b$ are empirical coefficients that relate the
opacity and the estimates of the atmosphere path delay,
$\tau_{\rm atm}(t)$ is the estimate of the non-hydrostatic atmosphere
path delay, $B^1_i(t)$ is the B-spline of the first degree with the
pivotal element $i$, and $T_{ri}$ are the coefficients of the expansion
the receiver temperature into B-spline basis. I set $T_{\rm atm}$ to
280K, and evaluated the coefficients of the receiver temperature
and regression parameters $a$ and $b$ by fitting them into
measurements of system temperatures with the use of the non-linear
LSQ. Parameter $a$ describes the possible bias between estimates
of the integrated water vapor contents and the non-hydrostatic path
delay in the zenith direction. The system temperature divided
by $e^{-(a + b\tau_{\rm atm}(t)) \, m(e)}$ is free from absorption
in the atmosphere and equivalent to that at the top of the atmosphere.

   The root mean squares (rms) of residuals of the empirical model
of the system temperature are in the range of 2--8~K.
The time variations of $T_{r}$ were within 10~K at {\sc eflsbereg},
{\sc jodrells2}, {\sc medicina}, and {\sc noto}. The $T_{r}$ was
around 180~K during first 3 hours at {\sc yebes40m} and then suddenly
dropped to 30~K and after that stayed stable. The $T_{r}$ at
{\sc onsala60} was unstable during the entire experiment and varied
in the range 40--250~K.

  The fringe amplitudes were calibrated by multiplying them by
the system temperature reduced to the top of the atmosphere and
dividing by the elevation dependent a~priori gain.

  The a~priori antenna gain and/or the term
$T_{\rm sys}/e^{(-a + b\tau_{\rm atm}(t)) \, m(e)}$ may have
a multiplicative error. The corrections to antenna gain were
evaluated by fitting the correlated amplitude to the flux density
of sources with known brightness distributions. Among sources used
as amplitude calibrators, brightness distributions are publicly
available\footnote{The database of brightness distributions,
correlated flux densities, and images of compact radio sources
produced with VLBI is accessible from
\web{http://astrogeo.org/vlbi\_images}} for 45 objects from
the K/Q survey and the VGaPS observing campaigns.

  For each used amplitude calibrator observation with known
brightness distribution in the form of CLEAN components I predicted
the correlated flux density $F$ as
\beq
     F_{corr} & = & \left|
                  \dss\sum_i c_i e^{\frac{2\pi\, f}{c}\, (u\, x + v\, y)}
                  \right| ,
\eeq{e:7}
   where $c_i$ is the correlated flux density of the $i$th CLEAN
component with coordinates $x$ and $y$ with respect to the center
of the image, and $u$ and $v$ are the projections of the baseline
vectors on the tangential plane of the source.

  Then I built a system of least square equations for all observations
of calibrators with known brightness distributions used in
the astrometric solution:
\beq
     F_{corr} = \sqrt{g_i \, g_j} A_{corr}
\eeq{e:8}
   and after taking logarithms from left and right hand sides solved
for corrections to gains $g$ for all stations using LSQ. Finally,
I applied corrections to gain for observations of all other sources.

   The gain correction was in the range 0.7--1.3 for all antennas,
except {\sc jodrell2}. The correction to {\sc jodrell2} gain was 6.9,
and its system equivalent flux density (SEFD) at elevations higher than
50\degr was around 7000~K.

   The detection limit varied significantly between the baselines.
Sources as weak as 13--16~mJy had the $\SNR>6.14$ and therefore, were
detected at the most sensitive baseline {\sc eflsberg}/{\sc yebes40m}.
The detection limit at baselines {\sc eflsberg}/{\sc medicina} and
{\sc eflsberg}/{\sc noto} was in the range of 25--30~mJy, at the baseline
{\sc eflsberg}/{\sc onsala60} was 30--40~mJy, and at the baselines
{\sc eflsberg}/{\sc jodrell2} and {\sc medicina}/{\sc noto} was
60--70~mJy.

\section{Data analysis: H${}_2$O maser sources}

\begin{figure}[h]
  \ifdraft
     \includegraphics[width=0.48\textwidth,clip]{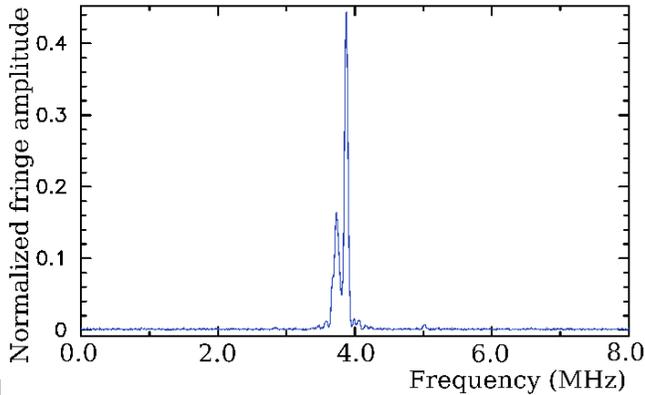}
  \else
     \includegraphics[width=0.48\textwidth,clip]{0221+618_medicina_onsala_1200.eps}
  \fi
  \caption{The normalized fringe amplitude of the water maser 0221$+$618
           at baseline {\sc medicina}/{\sc onsala60}. The horizontal
           axis shows the frequency offset with respect to the nominal
           H${}_2$O maser frequency ($6_{16}$--$5_{23}$) 22.235044~GHz.
  }
  \label{f:0221+618}
\end{figure}

  During scrutinizing the cross-correlation spectra, I found several
objects with very strong peaks in the IF9, for example object
0221$+$618 at baseline {\sc medicina}/{\sc onsala60}
(see Figure~\ref{f:0221+618}).

  This feature of the fringe spectrum is interpreted as an emission from
a water maser. The peak frequency of the cross-spectrum is shifted
within several megahertz due to a relative motion of the maser with
respect to the geocenter. Although this project was not targeted on
maser investigation, nevertheless, I decided to systematically search
for masers in the data, determine their coordinates and parameters
of their spectra.

   Processing the data from narrow-band sources differs significantly
from processing the data from continuum sources. I searched for masers
by running 31 trial fringe fitting with spectral window of 1~MHz within
the IF9 in the range of 22.22799--22.24399~GHz. Other spectral
constituents were masked out. The spectral window was shifted
at 0.5~MHz after each run. Since the maser has emission only within
a narrow band, suppressing the spectrum beyond the search window
improved the \SNR\ considerably, because the amount of noise is
reduced, but the power of the signal remained the same. I searched for
objects with the peak fringe amplitude higher than 4 times the spectrum
rms. This approach helped to discover 9 maser objects. After
the signal from masers was detected at at least one baseline,
the spectrum filter was narrowed down even further to reduce
the contribution of noise away from the spectral lines. This allowed
me to increase the number of detections at other baselines.

   The frequency resolution of the cross-spectrum, 125~KHz,
that correspond to the Doppler velocity resolution 1.7~km~s${}^{-1}$,
is not sufficient to resolve the water maser spectral line.
The scans with 7 out of 9 detected masers were re-correlated with 16 times
higher spectral resolution of 7.8~KHz, which corresponds to the
Doppler velocity resolution 0.1~km~s${}^{-1}$. Two other masers,
2130$+$556 and 2247$+$596 were discovered after the recorded data
were recycled, and it was too late to re-correlate them.

\begin{figure}[h]
  \ifdraft
     \includegraphics[width=0.48\textwidth,clip]{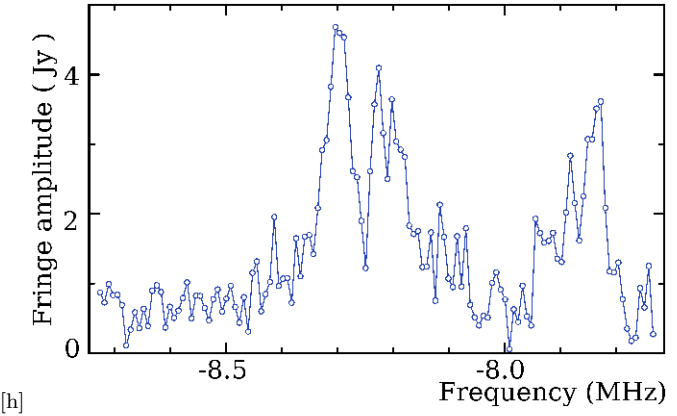}
  \else
     \includegraphics[width=0.48\textwidth,clip]{1923+151_eflsberg_medicina_1200.eps}
  \fi
  \caption{The fringe amplitude of the new water maser 1923$+$151
           at baseline {\sc eflsberg}/{\sc medicina}. The horizontal
           axis shows the frequency offset with respect to the nominal
           H${}_2$O maser frequency 22.235044~GHz.
  }
  \label{f:1923+151}
\end{figure}

\begin{figure}[h]
  \ifdraft
     \includegraphics[width=0.48\textwidth,clip]{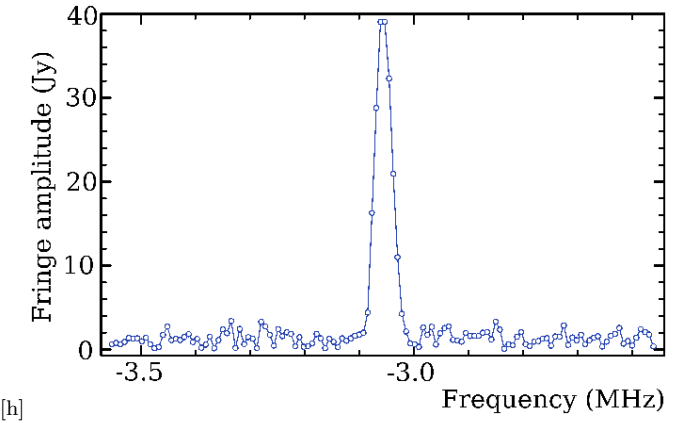}
  \else
     \includegraphics[width=0.48\textwidth,clip]{2011+360_eflsberg_jodrell2_1200.eps}
  \fi
  \caption{The fringe amplitude of the new water maser 2011$+$360
           at baseline {\sc eflsberg}/{\sc jodrell2}. The horizontal
           axis shows the frequency offset with respect to the nominal
           H${}_2$O maser frequency 22.235044~GHz.
  }
  \label{f:2011+360}
\end{figure}

   I smoothed the cross-spectrum and fitted the preliminary model
of the amplitude spectrum of maser emission
$A = \sum_i A_i \exp{-\{(f_i^s-f_0)^2/s_i^2\}}$
using an iterative non-linear LSQ procedure. Here $A_i$ is the peak
amplitude of the component, $f_i$ its frequency, $s_i$ is the
parameter of the line width, and $f_0$ is the rest H${}_2O$ frequency.
After determining a satisfactory preliminary model of the smoothed
spectrum, I fitted the final model using the raw cross-spectrum.
Since masers are not point-like objects, but usually a conglomerate
of tens or even hundreds spots, the correlated flux density
at different parts of the $uv$-plane differs considerably due
to beatings from different components. Using only 10--20 points at
the $uv$-plane is not sufficient for an image reconstruction.
Therefore, I restricted the amplitude analysis to determination of the
minimal and maximum correlated flux densities of detected components
at different baselines.

  The fringe fitting process cannot provide a meaningful estimate of
a group delay for a narrow-band source. Therefore, I resorted to using
phase delay rates for determining the source positions. A phase delay
rate over 2 minutes of integration time at 1024~Gb/s recording rate
has the uncertainty due to the thermal noise of order
$2{\cdot}10^{-13}/\SNR$. However, unmodeled short-term atmosphere
path delay fluctuations limit the accuracy of modeling the phase delay
rate usually at a level of 5--10${\cdot}10^{-14}$. The uncertainty of
the phase delay rate equivalent to group delay, $\sigma_r/\Omega_n$
(where at $\Omega_n$ is the nominal Earth rotation rate) \SNR=10 is 300~ps
if to take into account only thermal noise, and 1400~ps if to consider
the contribution of the atmosphere path delay rate. The uncertainty of
group delay at \SNR=10 in our experiment is around 200~ps, i.e. a factor
7~better. If to use only phase delay rates, the adjustments to nuisance
parameters, such as atmosphere path delays in zenith direction, station
positions, and clock functions, will be determined less precisely than
from the group delay solution and they will affect estimate of source
coordinates. In order to alleviate the effect of poorly determined
nuisance parameters on estimates of maser positions, I ran a special
solution with observation equations using both group delay and phase
delay rates for all detected sources, continuum spectrum objects and
masers. The phase delay rate had zero weight for continuum sources
and group delay had zero weight for maser sources.

  Positions of maser sources and their uncertainties are presented
in Table~\ref{t:masers}. This table also contains the peak minimum
and maximum correlated flux densities for each component of a maser
source over observations at different baselines and the velocity
with respect to the local system of rest (LSR) assuming the barycenter
of the Solar system moving with the speed of 20~km${}^{-1}$s towards
the direction with right ascension $18^h03^m50^s\!.24$ and declination
$+30\degr00'16''\!.8$. The table contains the full-width half-maximum
(FWHM) of the line profile and the minimum and maximum correlated flux
densities integrated over the line profile: $\int F(v)\ dv$. For two
sources that were correlated only with the low spectral resolution, the
low limit of their peak correlated flux densities and the low limit of
the FWHM of their line profiles are presented.

\begin{table*}
  \caption{Properties of detected H${}_2$O masers. Column description:
           (1) IAU name, (2) common name, (3) IRAS designator
           (4) right ascension, (5) uncertainty in right ascension
               in arcsec (without $\cos\delta$ factor),
               (6) declination, (7) uncertainty in
               declination in arcsec, (8) component index,
               (9) minimum correlated flux density in Jy,
               (10) maximum correlated flux density in Jy,
               (11) LSR velocity in km${}^{-1}$s,
               (12) FWHM of the line profile in km${}^{-1}$s,
               (13) minimum integrated flux density in Jy km${}^{-1}$s,
               (14) maximum integrated flux density in Jy km${}^{-1}$s.
          }
  \begin{tabular}{l l l r r r r r r r r r r r }
         \hline
         \ntab{c}{(1)}  & \ntab{c}{(2)}  & \ntab{c}{(3)}  &
         \ntab{c}{(4)}  & \ntab{c}{(5)}  & \ntab{c}{(6)}  &
         \ntab{c}{(7)}  & \ntab{c}{(8)}  & \ntab{c}{(9)}  &
         \ntab{c}{(10)} & \ntab{c}{(11)} & \ntab{c}{(12)} &
         \ntab{c}{(13)} & \ntab{c}{(14)}
         \vex \\
         \hline
         0221$+$618  & W3(2)  & 02219+6152 & 02\ 25\ 40.672 & 0.13 & +62\ 05\ 54.06 & 0.08
                     &  1 &  24.6 &  52.1 & $    -59.0$ &  0.4 &   5.7 &  12.1 \\
         & & & & & & &  2 &   6.0 &  28.3 & $    -58.5$ &  0.3 &   1.1 &   5.9 \\
         & & & & & & &  3 &  13.7 & 1039.1 & $   -55.9$ &  0.4 &   3.3 & 275.5 \\
         & & & & & & &  4 &   4.4 & 393.4 & $    -54.3$ &  0.6 &   0.8 & 120.3 \\
         & & & & & & &  5 &   5.6 & 324.0 & $    -53.6$ &  0.5 &   0.9 &  87.7 \\
         & & & & & & &  6 &  15.8 & 346.7 & $    -52.3$ &  0.5 &   4.5 & 102.5 \\
         & & & & & & &  7 &  12.1 &  85.7 & $    -51.0$ &  0.4 &   2.5 &  26.1 \\
         & & & & & & &  8 &   7.4 &  42.6 & $    -49.7$ &  0.4 &   1.5 &  11.4%
         \vex\vex \\
         1907$+$082  &        & 19074+0814 & 19\ 09\ 49.847 & 0.03 & +08\ 19\ 45.28 & 0.20
                     &  1 &   7.5 &  19.4 & $\hm 104.6$ &  0.5 &   2.1 &   5.0 \\
         & & & & & & &  2 &   1.8 &   1.8 & $\hm 112.0$ &  0.4 &   0.4 &   0.4%
         \vex\vex \\
         1920$+$143  &  W51 W & 19209+1421 & 19\ 23\ 11.224 & 0.03 & +14\ 26\ 45.80 & 0.11
                     &  1 &   6.9 &   9.0 & $\hm  96.5$ &  0.6 &   1.8 &   3.8 \\
         & & & & & & &  2 &   2.8 &  26.5 & $\hm  98.6$ &  0.5 &   0.7 &   7.4 \\
         & & & & & & &  3 &   3.1 &   9.9 & $\hm 100.1$ &  0.5 &   0.7 &   3.8 \\
         & & & & & & &  4 &   8.8 &  34.2 & $\hm 101.9$ &  0.5 &   2.7 &  10.2 \\
         & & & & & & &  5 &   1.9 &   5.1 & $\hm 103.3$ &  0.4 &   0.4 &   1.4%
         \vex\vex \\
         1923$+$151  &        & 19230+1506 & 19\ 25\ 17.907 & 0.06 & +15\ 12\ 24.46 & 0.17
                     &  1 &   1.6 &   1.8 & $\hm  53.0$ &  0.5 &   0.4 &   0.5 \\
         & & & & & & &  2 &   5.3 &   5.3 & $\hm  55.0$ &  0.8 &   2.4 &   2.4 \\
         & & & & & & &  3 &   1.7 &   4.1 & $\hm  57.4$ &  0.4 &   0.5 &   0.8 \\
         & & & & & & &  4 &   2.6 &   3.6 & $\hm  58.8$ &  0.4 &   0.6 &   0.8%
         \vex\vex \\
         2011$+$360  &        & 20116+3605 & 20\ 13\ 34.296 & 0.06 & +36\ 14\ 54.36 & 0.06
                     &  1 &   8.9 &  45.4 & $    -19.1$ &  0.3 &   1.6 &   7.9%
         \vex\vex \\
         2107$+$521  &  WB43  & X2107+521  & 21\ 09\ 21.717 & 0.04 & +52\ 22\ 37.05 & 0.03
                     &  1 &   1.5 &  55.6 & $     -3.5$ &  0.4 &   0.3 &  16.0 \\
         & & & & & & &  2 &   2.2 &  41.0 & $\hm   2.4$ &  0.4 &   0.5 &  16.6 \\
         & & & & & & &  3 &   2.5 &  63.4 & $\hm   4.1$ &  0.6 &   0.7 &  26.1 \\
         & & & & & & &  4 &   6.5 &  17.3 & $\hm   8.7$ &  0.4 &   1.3 &   3.8 \\
         & & & & & & &  5 &   5.3 &  25.1 & $\hm  11.8$ &  0.4 &   1.1 &   6.5 \\
         & & & & & & &  6 &   2.3 &  40.1 & $\hm  12.8$ &  0.4 &   0.6 &   7.5 \\
         & & & & & & &  7 &   1.6 &  37.0 & $\hm  13.8$ &  0.3 &   0.3 &   7.7 \\
         & & & & & & &  8 &   1.3 &  38.1 & $\hm  14.8$ &  0.3 &   0.3 &   7.5 \\
         & & & & & & &  9 &   2.3 &  99.9 & $\hm  16.2$ &  0.3 &   0.4 &  20.3 \\
         & & & & & & & 10 &   2.8 &  16.1 & $\hm  17.4$ &  0.3 &   0.6 &   2.8%
         \vex\vex \\
         2130$+$556  &        & 21306+5540 & 21\ 32\ 12.444 & 0.07 & +55\ 53\ 49.64 & 0.04
                     &  1 &  0.8 & $>32.3$ & $    -60.9$ & $<1.7$ &  0.6 &  25.3%
         \vex\vex \\
         2247$+$596  &  S146  & X2247+596  & 22\ 49\ 31.474 & 0.06 & +59\ 55\ 41.91 & 0.03
                     &  1 &   1.1 & $>17.2$ & $    -47.4$ & $<1.7$ &   0.8 &  11.6%
         \vex\vex \\
         2254$+$617  &  Cep A & 22543+6145 & 22\ 56\ 17.988 & 0.06 & +62\ 01\ 49.40 & 0.04
                     &  1 &   2.7 &  34.8 & $    -34.2$ &  0.4 &   0.5 &   7.7 \\
         & & & & & & &  2 &   3.2 &  31.4 & $    -22.6$ &  0.5 &   0.6 &  10.7 \\
         & & & & & & &  3 &   2.9 &  37.4 & $    -16.0$ &  0.6 &   0.5 &  15.6 \\
         & & & & & & &  4 &  11.7 &  20.6 & $    -14.7$ &  0.4 &   2.3 &   4.3 \\
         & & & & & & &  5 &   4.5 &   5.2 & $     -9.9$ &  0.3 &   0.9 &   0.9 \\
         & & & & & & &  6 &  21.7 &  40.1 & $     -7.2$ &  0.4 &   5.0 &  11.8 \\
         & & & & & & &  7 &   5.7 & 111.8 & $     -5.9$ &  0.5 &   1.3 &  47.9 \\
         & & & & & & &  8 &   3.1 & 217.7 & $     -4.4$ &  0.3 &   0.6 &  36.8 \\
         & & & & & & &  9 &   5.0 &   8.4 & $     -2.9$ &  0.5 &   1.3 &   1.9 \\
         & & & & & & & 10 &   4.9 &   5.4 & $\hm   0.1$ &  0.3 &   1.0 &   1.1 \\
         & & & & & & & 11 &   4.3 &   7.5 & $\hm   1.8$ &  0.5 &   0.8 &   2.8%
         \\
         \hline
  \end{tabular}
  \label{t:masers}
\end{table*}

\subsection{Comments on individual masers}

{\bf Cep A}. \citet{r:cur02} reported detection of
a 2~mJy continuum source at 7~mm in the \mbox{Cep A} complex
from VLA observations in 1996. Position of that source,
which they labeled as VLA-mm, is $22^h56^m17^s\!.985$
$+62\degr01'49''\!.45$ with the uncertainty $0''.02$.
Our position of \mbox{Cep A} water maser coincides
within $1\sigma$ with the position of VLA-mm source.
\citet{r:cepa} determined position of the methanol
maser associated with this object using phase-referencing VLBI:
$22^h56^m18^s.0970$  and $62{\degr}01'49''\!.399$ with
sub-milliarcsecond accuracy. Their position of the methanol
maser is within $0''\!.78$ from the water maser.

{\bf \object{1923$+$151}} is a new discovery. Its spectrum is shown in
Figure~\ref{f:1923+151}. It is associated with
\object{IRAS19230$+$1506} ($3''\!.8$ away) and WISE
\object{J192517.93$+$151225.0}  ($0''\!.8$ away), within
1--2$\sigma$ of position uncertainty of these catalogues.

{\bf \object{2011$+$360}} is a new discovery. Its spectrum is shown
in Figure~\ref{f:2011+360}. It is associated with
\object{IRAS20116$+$3605} ($8''\!.6$ away). The object was
selected by \citet{r:sunada} as a water maser candidate, was
observed, but not detected.

\section{The EGaPS catalogue}
\label{s:catalog}

\begin{table*}
   \caption{The first 12 rows of the EGaPS catalogue of source positions
            of 109 target sources. The table columns are explained
            in the text. The full table is available in the electronic
            attachment.}
   \label{t:lcs1}
   \footnotesize
   \begin{tabular}{ c l l l r r r r r r r r r l l}
      \hline
         Class    &
         IVS      &
         IAU      &
         Right $\,$ ascension &
         Declination     &
         $\Delta \alpha$ &
         $\Delta \delta$ &
         Corr   &
         \# Obs &
         Total  &
         Unres
         \vspace{0.5ex} \\
         \ntab{c}{(1)}    &
         \ntab{c}{(2)}    &
         \ntab{c}{(3)}    &
         \ntab{c}{(4)}    &
         \ntab{c}{(5)}    &
         \ntab{c}{(6)}    &
         \ntab{c}{(7)}    &
         \ntab{c}{(8)}    &
         \ntab{r}{(9)}    &
         \ntab{c}{(10)}   &
         \ntab{c}{(11)}
         \vspace{0.5ex} \\
U  & 0002$+$576 & J0004$+$5754 & 00 04 50.263646 & $+$57 54 57.75926 & 16.9 & 18.9 & -0.159 &  4 & $ 0.022 $ & $ -1.000 $ & \vspace{0.5ex} \\
N  & 0008$+$657 & J0011$+$6603 & 00 11 38.823302 & $+$66 03 38.51426 & 27.0 &  4.9 &  0.245 &  7 & $ 0.043 $ & $ -1.000 $ & \vspace{0.5ex} \\
C  & 0017$+$627 & J0019$+$6300 & 00 19 47.644959 & $+$63 00 46.63373 & 28.8 &  4.9 &  0.094 &  8 & $ 0.037 $ & $  0.047 $ & \vspace{0.5ex} \\
N  & 0133$+$602 & J0136$+$6032 & 01 36 56.806879 & $+$60 32 05.06934 & 24.7 &  6.2 & -0.159 &  7 & $ 0.035 $ & $ -1.000 $ & \vspace{0.5ex} \\
U  & 0158$+$623 & J0201$+$6237 & 02 01 52.980387 & $+$62 37 57.12279 & 48.4 & 11.0 & -0.305 &  4 & $ 0.025 $ & $ -1.000 $ & \vspace{0.5ex} \\
N  & 0245$+$620 & J0248$+$6214 & 02 48 58.890890 & $+$62 14 09.66282 & 19.8 &  5.8 & -0.296 &  6 & $ 0.057 $ & $ -1.000 $ & \vspace{0.5ex} \\
N  & 0252$+$574 & J0256$+$5736 & 02 56 29.004510 & $+$57 36 42.45601 & 12.9 &  4.5 & -0.013 &  9 & $ 0.070 $ & $ -1.000 $ & \vspace{0.5ex} \\
U  & 0313$+$531 & J0317$+$5318 & 03 17 01.346538 & $+$53 18 27.51902 & 19.8 &  9.8 &  0.225 &  4 & $ 0.032 $ & $ -1.000 $ & \vspace{0.5ex} \\
C  & 0334$+$565 & J0338$+$5640 & 03 38 36.909254 & $+$56 40 50.05381 & 11.9 &  3.1 &  0.034 & 11 & $ 0.071 $ & $  0.078 $ & \vspace{0.5ex} \\
N  & 0339$+$573 & J0343$+$5729 & 03 43 44.815127 & $+$57 29 08.52764 & 26.0 &  8.6 &  0.174 &  5 & $ 0.034 $ & $ -1.000 $ & \vspace{0.5ex} \\
N  & 0357$+$457 & J0400$+$4554 & 04 00 34.574245 & $+$45 54 24.05505 & 11.1 &  5.4 &  0.204 &  7 & $ 0.068 $ & $ -1.000 $ & \vspace{0.5ex} \\
      \hline
      \hline
   \end{tabular}
   \label{t:egaps}
\end{table*}

  Table~\ref{t:egaps} displays 12 out of 109 rows of the EGaPS
catalogue of source positions. The full table is available in the
electronic attachment. Column 1 contains the class of the source,
columns 2 and 3 contain the J2000 and B1950 IAU names; column 4,
and 5 contain right ascensions and declinations of sources. Columns
6 and 7 contain position errors in right ascension (without
multiplier $\cos \delta$) with noise 4.0~mas/$\cos \delta$ and
1.0 mas added in quadrature to formal position uncertainties
in right ascension and in declination respectively. Column 8 contains
correlation coefficients between right ascension and declination,
column 9 contains the total number of observations used in position
data analysis. Columns 10 and 11 contain the median estimates of the
correlated flux density over all experiment of the source in two
ranges of baseline projection lengths: 0--67~$M\lambda$ (projections
shorter than 900~km) and 100-170~$M\lambda$ --- projections longer
than 1350~km. If there were no scheduled observations at the range
of the baseline projection lengths, then $-1.000$ is put in the
table cell.

\section{Summary}
\label{s:summary}

  I determine positions of 109 compact extragalactic sources that
are within 6\degr of the Galactic plane with declinations $> -20\degr$.
Observations were made at a six-element array of the Europearn VLBI Network.
As a result of these observations, the total number of
calibrator sources with precisely known positions in this region grew
from 322 to 431. The semi-major axis of position errors range from
5 to 60~mas, and the median position uncertainty is 9~mas.

  Among detected sources, nine water masers in our Galaxy were identified,
two of them new. The date for seven these objects were re-correlated with
the spectral resolution of 0.1 km/s. I determined position of masers from
delay rates with accuracies 30--200~mas, determined velocities of components
and their correlated flux densities.

  I also obtained estimates of source correlated flux densities.
The correlated flux densities of extragalactic sources range from
20 to 300~mJy with the median value of 55~mJy.

  The results of this campaign are somewhat disappointing. First,
the median position accuracy is one order of magnitude worse than
in a similar campaign with the VLBA. Several factors contributed to
the increase of uncertainties in source positions. First,
the unfavorable frequency setup resulted in an increase of group delay
uncertainties by a factor of 2.4. Retrospectively, I should admit that
more efforts should be put in order to resolve logistical issues with
LO frequency changes. Second, the size of the EVN network used in
this experiment is a factor of 3 less than the size of the VLBA
network. Third, the sources in EGaPS campaign  are a factor of
3~weaker than in the VGaPS campaign.

  The detection rate, 30\% is significantly lower than in other
survey experiments. For comparison, the detection rate in the VGaPS
experiment among sources in the Galactic plane preselected with VERA
observations was 76\% and 36\% among the sources selected on the basis
of their spectral index. Apparently, the efficiency of sources
selection based on the spectral index is at a level of 30-40\% in
the Galactic plane. About 1/3 detected sources have the correlated
flux density below 40~mJy which makes their use as phase calibrator
problematic. This indicates that the source selection strategy
should be revised if a similar survey will be made in the future.

\section{Acknowledgments}
\label{s:acknowledgments}

I made use of the database CATS of the Special Astrophysical
Observatory. I used in our work the dataset MAI6NPANA provided by the
NASA/Global Modeling and Assimilation Office (GMAO) in the framework
of the MERRA atmospheric reanalysis project. It is my pleasure to
thank Yury Kovalev, David Graham, Richard Porcas, and Cormac Reynolds
for useful suggestions that contributed to the success of the experiment.
I would like to thank Alessandra Bertarini and Helge Rottmann for
prompt re-correlation scans with masers. The European VLBI Network
is a joint facility of European, Chinese, South African, Russian,
and other radio astronomy institutes funded by their national
research councils. This publication makes use of data products from
the Wide-field Infrared Survey Explorer, which is a joint project
of the University of California, Los Angeles, and the Jet Propulsion
Laboratory/California Institute of Technology, funded by the NASA.

\label{lastpage}

\end{document}